\documentclass[journal,12pt,onecolumn,letterpaper]{IEEEtran}
\usepackage{arxiv}
\usepackage{geometry}
\usepackage{times}
\usepackage{cite}
\usepackage{url}
\usepackage{graphicx}
\usepackage{lscape}
\usepackage{subcaption}
\usepackage{rotating}
\usepackage{rotfloat}
\usepackage{xcolor}
\usepackage{amsmath}
\usepackage{amssymb}
\usepackage{algorithm}
\usepackage{algpseudocode}
\usepackage{array}
\usepackage[english]{babel}
\usepackage{gensymb}
\usepackage{textcomp}
\usepackage{placeins}
\usepackage{balance}
\usepackage{booktabs}

\title{Beyond the Trust Boundary: A Critical Reassessment of the FIDO2 Threat Model}

\author{
 
    Aditya Mitra \\
   CyberMACS\\
   Kadir Has University \\
    DigitalFortress Private Limited  \&\\
    Indominus Labs Private Limited \\
    \texttt{adityaarghya0@gmail.com}

\And
Kolluru Sai Abhiram \\
     Centre of Excellence, Cyber Security \\
   School of Computer Science and Engineering \\
VIT-AP University, India \\
    DigitalFortress Private Limited \\
\And
  Sibi Chakkaravarthy Sethuraman\\
    Centre of Excellence, Artificial Intelligence \& Robotics (AIR),\\
    School of Computer Science and Engineering\\
    VIT-AP University, India \\
     DigitalFortress Private Limited  \&\\
    Indominus Labs Private Limited \\
    \texttt{sb.sibi@gmail.com} \\

\And

 Anitha S \\
     Centre of Excellence, Artificial Intelligence and Robotics (AIR)  \\
   School of Electronics and Communication Engineering \\
VIT-AP University, India \\
    DigitalFortress Private Limited \\
}

\begin{document}
\maketitle

\begin{abstract}
FIDO2/WebAuthn has been deployed extensively as a phishing resistant authentication scheme. Since FIDO2 uses public-key cryptography and requires the presence of a hardware backed authenticator, its security has long been considered guaranteed by its design (assuming its cryptographic implementation is correct), both from the perspective of practitioners and prior academic work. In this work, we dissect the FIDO2 threat model to show that some of the commonly assumed security properties do not hold in all circumstances, and we also extend the threat model to cover more than just the cryptographic layer, as was previously done in works such as \cite{dbase:fido2-formal-analysis,bindel2022ctap21,localattacks2023}. We look at eight potential attack vectors associated with various layers of the FIDO2 stack --- malicious web extensions, platform handler malware, passive sniffing, virtual device drivers, CTAP2 specific malware, USB/hardware implants, malicious USB hubs/docks/extenders, and NFC relay attacks --- and show how FIDO2 relies on certain assumptions about the environment which are frequently not true in practice. We show how AAGUID and timing information can be used to profile users and carry out targeted attacks \cite{mojoauth2026aaguid,yubico_attestation} and show that compromising the environment at any layer --- the browser, the OS, or the hardware --- can affect the security of FIDO2. Furthermore, we show how some of these attacks can be chained together to undermine the security of FIDO2 even when the underlying cryptographic primitives are implemented correctly. These findings support the thesis that the weak link in a FIDO2 deployment is rarely the crypto layer. We also show how several of these attack vectors can undermine the device attestation model that underlies FIDO2, by subverting the FIDO Metadata Service (MDS3) that serves as its root of trust \cite{fidomds2021}, and we characterize each of these attacks by their privilege, skill, and resource requirements. We conclude that the mitigations for these attacks need to be layered and address problems at all levels of the FIDO2 stack, protect the metadata, and ensure that the environment is protected.

\end{abstract}


\section{Introduction}
\label{sec:introduction}

FIDO2/WebAuthn (hereby WebAuthn) is a standard intended to replace passwords with a better, phishing resistant authentication scheme based on public key cryptography. The WebAuthn standard was defined and standardized by the W3C and the FIDO Alliance~\cite{w3cwebauthn2021,fido2vectors2026}. FIDO2 is made up of two separate but complementary specifications: the Web Authentication (WebAuthn) standard, which defines a JavaScript API for creating and utilizing strong attested public key credentials, and the Client to Authenticator Protocol (CTAP) specification, which specifies the protocol that is used to facilitate communication between a browser or client and external authenticators~\cite{fidoctap2}. It is important to note that CTAP is a distinct, standalone FIDO Alliance specification and not a component of the W3C WebAuthn standard itself; the two specifications are designed to interoperate but are maintained and versioned independently.

The FIDO2 protocol has gained popularity over the last few years among various platforms and services. Apple was the first major platform vendor to ship support for FIDO2, followed by Microsoft and Google adding support for FIDO2 to Windows and Android, respectively. Many web services also now support FIDO2.

FIDO2 offers a possible defense against phishing, man-in-the-middle, and credential stuffing attacks that affect passwords. Through its reliance on public-key cryptography and physical user interaction, FIDO2 could theoretically eliminate the risk of attackers being able to capture and replay secrets exchanged over insecure channels. Naturally, as with any technology, security relies heavily on the assumptions made about the environment and threat model, and is only as strong as its weakest component. When those assumptions fail, or can be circumvented through adversarial actions, the security guarantees of FIDO2 are violated~\cite{localattacks2023,fido_security_ref}.

In this paper, we analyze the threat model upon which FIDO2 was designed, decompose it into its underlying assumptions, and identify different scenarios in which this threat model can break down. Furthermore, we show how the attacks we identify can be combined to form complex exploit chains. Our contributions are summarized as follows, A systematic breakdown of the FIDO2 threat model into its assumptions and trust boundaries, a study of eight types of attacks that target different components of the FIDO2 stack, discussion of how the attacks could be chained together to produce stronger attacks and an assessment of the security impact of our findings on metadata privacy, environmental security, and the eroding of trust with Practical mitigation strategies and a methodology for studying the security of authentication systems by going beyond cryptographic properties and examining the security of the overall execution environment.

The rest of the paper is organized as follows. In Section\ref{sec:background} we give a brief overview of FIDO2/WebAuthn architecture and its relevant components. We present our refined threat model and discuss the implications of that threat model in Section\ref{sec:contribution}. In Section\ref{sec:attack-vectors}, we detail eight attack vectors targeting different components of the system. We discuss several kill chains that can be formed by combining these vectors in Section\ref{sec:kill-chains}. In Section\ref{sec:implications} we discuss how our work goes beyond the attack vectors, including how the leakiness of metadata can impact the security of the ecosystem, and how our work undermines core trust models underpinning the FIDO2 protocol. Section\ref{sec:results} outlines our experimental results and how we analyzed them. We analyze each of our attacks in terms of required attacker capabilities and resources in Section\ref{sec:attacker-capabilities}. Finally, we conclude in Section\ref{sec:conclusion} with a summary of our work and possible future directions.

\noindent{\textit{Note on scope:}} To avoid the risk of our work being directly deployed as attack tooling, the attack sections contain only a description of the mechanism, demonstrated impact, and illustrative pseudocode at an architectural level. No code to fully implement the attack is presented. Where appropriate, links to existing publicly available tools that can be used alongside our work are provided. Any details that were reported to vendors that would be useful in isolation of our work have either been obfuscated (to maintain user privacy) or omitted entirely (as they are only useful in the context of the specific vulnerability).

\section{Background}
\label{sec:background}

To grasp the attacks described in this paper, we first need to understand the FIDO2 / WebAuthn architecture and trust model. This section provides a brief overview of these topics.

\subsection{Architecture}

FIDO2 comprises of two primary standardized specifications that work together to provide authentication services:

Web Authentication (WebAuthn): A W3C standard that defines an API for web applications to create and use strong, attested, scoped, public-key based credentials for the purpose of strongly authenticating users~\cite{w3cwebauthn2021}.

Client to Authenticator Protocol (CTAP): A FIDO Alliance standard defining a communication protocol between a client (typically a browser or OS) and an authenticator (a device that holds keys and performs authentication operations)~\cite{fidoctap2}. CTAP1/U2F is the original Universal Second Factor specification, designed for second factor authentication. CTAP2 is a newer iteration of CTAP designed to enable multi-factor authentication and additional features like resident keys and user verification.

Authenticators: Authenticators are devices that store credentials and perform cryptography. They can be \textit{Platform Authenticator} (built into a device like Windows Hello or Touch ID on Mac/iOS), or \textit{Cross-Platform Authenticators} (standalone devices that can move between systems, like USB security keys, NFC tokens, Smart Cards and Bluetooth authenticators).

\subsection{Trust Model and Assumptions}

The FIDO2 specification makes several implicit security assumptions~\cite{fido_security_ref}. The cryptographic scheme used to verify signatures (typically ECDSA over elliptic curves) is unbreakable under current computational capabilities (\textit{cryptographic assumption})~\cite{dbase:fido2-formal-analysis,bindel2022ctap21,fido2vectors2026}. The hardware security features of the authenticator successfully prevent attackers from extracting the authenticator's private key (\textit{hardware security}). The authenticator accurately detects when a user interacts with it by placing their finger or pressing a button (\textit{user presence}). The client's operating system is uncompromised by malware or other security breaches (\textit{trusted client}). The cryptographic scheme for verifying signatures (usually ECDSA on elliptic curves) is not computationally feasible to break (\textit{cryptographic assumption})~\cite{dbase:fido2-formal-analysis,bindel2022ctap21}. Authenticator hardware security ensures that the attacker cannot extract the authenticator's private key (\textit{hardware security}). Several implicit security assumptions are made in the FIDO2 specification. The cryptographic scheme for verifying signatures (usually ECDSA on elliptic curves) is not computationally feasible to break (\textit{cryptographic assumption})~\cite{dbase:fido2-formal-analysis,bindel2022ctap21}. Authenticator hardware security ensures that the attacker cannot extract the authenticator's private key (\textit{hardware security}). The authenticator reliably detects when the user places a finger or presses a button (\textit{user presence}). (SA-4, trusted client) The client OS is not compromised by a malicious program or any other security vulnerability. The link between client and authenticator is secure against eavesdropping and tampering (\textit{secure channel}). (SA-6, browser integrity \& server security) The web browser securely exposes the WebAuthn API to websites and acts as intermediary between the website and the underlying platform~\cite{squarex2025passkeyspwned}, and the server of the Relying Party correctly interprets responses to authentication requests, validates the value of signatures, and stores public keys in a manner where they cannot be modified without authorization.

All client components (the browser IPC layer, drivers, and the platform handler) depend on one another for trust without any further verification. This implicit trust is an attack surface, as we demonstrate in subsequent sections~\cite{localattacks2023}.

\subsection{Working Procedure}

There are two ceremonies defined by FIDO2, i.e. the registration ceremony where users create credentials and the authentication ceremony where users make assertions. For creating a credential, the user registers an account with the relying party, i.e. a website or a service. The user tries to sign up for the service and the server first issues a challenge that specifies what type of credentials the server expects from the browser (e.g. what type of authenticator, whether the user needs to provide a PIN code, etc.). Then the client contacts the authenticator that supports this challenge, and the user must authorize access by presenting proof of their presence (e.g. clicking a button on the authenticator device or using some biometric).

The authenticator then generates a new public-private key-pair for this particular origin (domain or service name) and saves the private key in a secure storage (a secure element), creates the public key and an attestation for the authenticator and sends them to the relying party's server. The server associates the public key with the user's account and saves it for later use in authentication. According to previous studies, FIDO2 can still have a good usability when it is implemented correctly~\cite{dbase:fido2-usability,ravilla2024study}.

For authentication, the user tries to log in to a service. The relying party's server generates a new challenge and sends it to the browser. The browser then requests an assertion for a credential tied to the origin of the relying party from the authenticator. Again, the user needs to give consent and proof of presence. The authenticator signs the challenge with other context information (e.g. origin and counter) and returns the signature to the browser. The browser sends the signature to the relying party's server, which verifies it with the public key saved during the registration phase. If the signature is valid and other conditions are met (e.g. the origin is correct and the counter is incremented), the authentication is completed successfully.

\section{Contributions}
\label{sec:contribution}

In our opinion, this paper is a valuable critique of FIDO2/WebAuthn security, as it highlights and catalogs a number of significant problems related to the implicit trust assumptions that are part of the FIDO2 standard’s threat model~\cite{fido2vectors2026,fido_security_ref}. Even though we offer no new cryptographic or implementation solutions, our primary contribution here is that we refine the conceptual framework by which the security of FIDO2 deployments is evaluated.

\begin{figure*}[t]
\centering
\includegraphics[width=\linewidth]{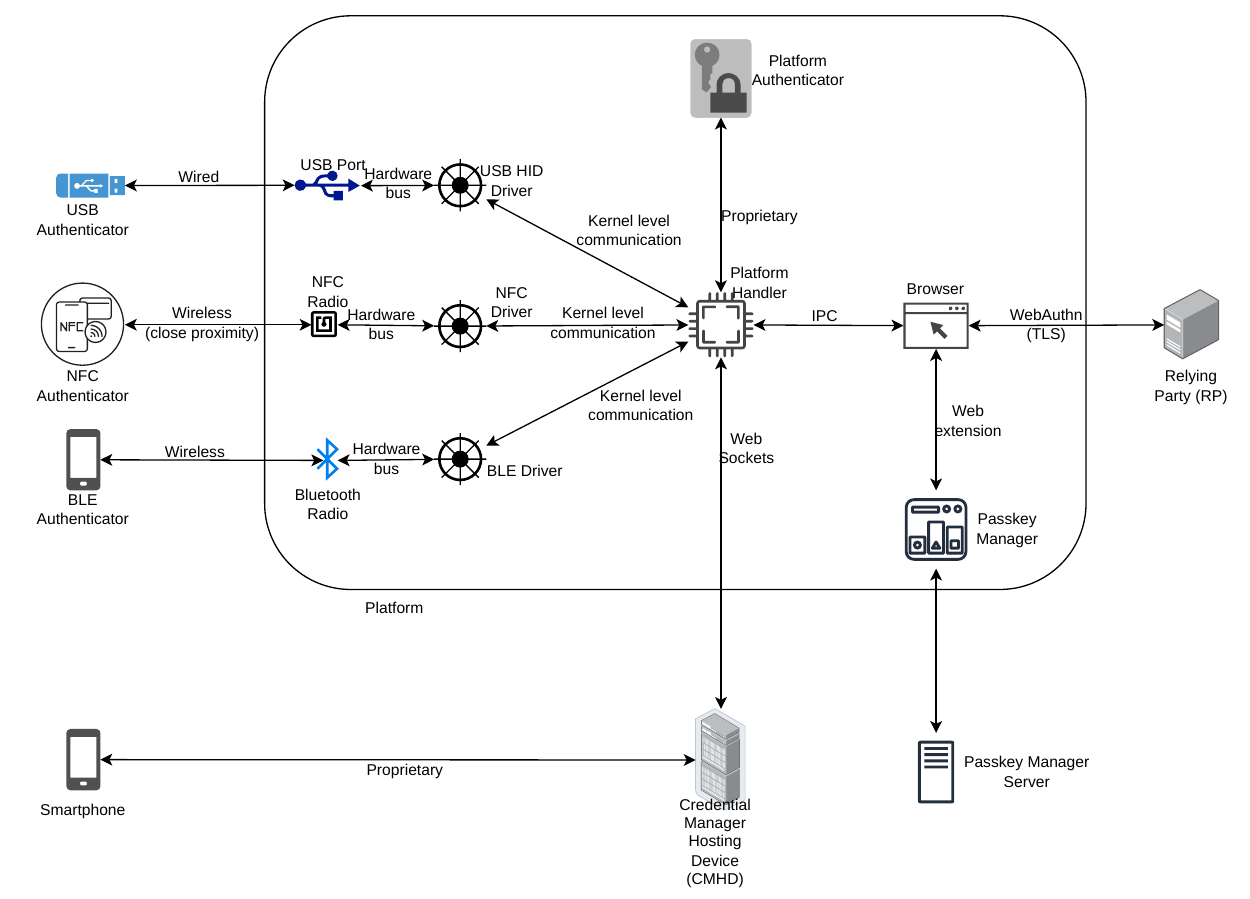}
\caption{Threat Model Diagram illustrating FIDO2/WebAuthn architecture and trust boundaries}
\label{fig:threat}
\end{figure*}

\subsection{Limitations of the Original Threat Model}
\label{subsec:original-model-limitations}

The standard FIDO2 threat model used by the standards documents and security literature, contains some routinely violated by attacker capabilities~\cite{fido_security_ref}. It analyses the threat model for each attack vector independently, while ignoring the fact that multiple attack vectors may be chained together for a stronger attack. It assumes the client (browser/OS/hardware) to be trusted, or at least, not fully compromised. It assumes attackers have limited capabilities, and doesn't analyze the threat model for all types of attackers, including malware authors and nation states. It assumes the trade boundary is fixed, while in reality it is dynamic. And finally, it focuses too much on the cryptographic primitives, and does not place sufficient emphasis on implementation and environmental security~\cite{dbase:fido2-enterprise-challenges}.

\subsection{Our Refined Threat Model}
\label{subsec:refined-model}

Our enhanced threat model addresses these shortcomings by representing attacks as chains, rather than individual attacks, and thereby reflecting the possibility that multiple individually-insignificant breaches could occur in sequence, each of which is individually not enough to bypass FIDO2’s protections but, together, allow the adversary to succeed. It also incorporates the concept of “environmental trust,” which acknowledges that an authenticator is not an isolated entity, but is part of a complex ecosystem of components, including device drivers, OS APIs, inter-process communication mechanisms, and external ports that mediate its connection to the client—and, by extension, to the RP. We then extend the threat model to note that the security provided by FIDO2 is dependent not only on the security of its cryptographic protocol, but on the security of its execution environment. Even if FIDO2 uses a perfectly secure signature scheme, if the environment in which it operates is insecure, its security guarantees will not hold. The threat model also covers a spectrum of attacker sophistication, from low-skill to state-level, enabling reasoning about a range of attacks that include opportunistic attacks (e.g., phishing, social engineering), as well as advanced attacks against attackers with unlimited resources.

It represents trust boundaries as continuously observable rather than static (that is, they were previously assumed to be trusted and never re-evaluated), where a trust boundary assumed to be secure might be compromised later (e.g., due to compromised firmware, supply-chain compromise, or privilege escalation). Our threat model also evaluates the security of FIDO2 at the whole-system level (as opposed to only the cryptographic level) by considering the security of all elements of the FIDO2 architecture: hardware, firmware, software, and users. Thus, the impact of security vulnerabilities in any one component of the FIDO2 stack, or combinations thereof, can be evaluated.

\subsection{Practical Implications of the Refined Model}
\label{subsec:practical-implications}

FIDO2-capable organizations (and those considering FIDO2) must now recognize that the scope of defense-in-depth extends beyond the authenticator, because the client, browser, and OS may also be compromised and undermine authentication guarantees, even when the authenticator itself is uncompromised. Trust property violation monitoring (e.g., detecting unauthorized USB devices or unusual authentication patterns) should not be a one-time assessment, but an ongoing control, because trust boundaries can degrade over time, e.g., due to firmware updates, device transfers, or privilege escalation on the host. FIDO2 cannot be relied upon as a sole protection against credential theft; network monitoring, endpoint detection, and user education should be applied in tandem with it, since FIDO2 is not intended to guard against a compromised execution environment.
Adaptive authentication (e.g., based on device health, location, or behavior) may offer an additional mitigation if the environment is compromised, because relying parties could be allowed to mark technically-valid assertions as suspicious if they come from an unhealthy or unexpected context (e.g., from a compromised device or one at an unexpected physical location). Authenticator vendors should be clearer about what security assumptions they make about the environment in which the authenticator will operate~\cite{dbase:fido2-enterprise-challenges}, so that organizations do not discover a gap in compensating controls only after a breach occurs. Ultimately, clarifying the threat model enables security architects, developers, and administrators to stop thinking of FIDO2 as an unconditional guarantee of phishing-resistance, and instead begin thinking about the actual security properties (which depend on the environment) that it delivers.

\section{Related Work}
\label{sec:related-work}

FIDO2/WebAuthn security has previously been studied through the lens of formal protocol verification, local and implementation attacks, metadata and side-channel leakage, and usability/enterprise-deployment issues. We build on these lines of work, extending them by moving from studying FIDO2 components to studying the entire system.

\subsection{Formal and Cryptographic Analysis}

Much previous work has verified the cryptographic core of FIDO2 and CTAP formally~\cite{dbase:fido2-formal-analysis}. The authors define an idealized protocol, then show that an attacker who does not possess the authenticator private key cannot forge a signature. Subsequent work has extended this model to CTAP2.1, specifically verifying the security of the new PIN and user-verification sub-protocols~\cite{bindel2022ctap21}. While useful for establishing the correctness of the protocol itself, these analyses typically assume a trusted client, browser, and transport, assumptions that we explicitly relax in this work.

\subsection{Local and Implementation-Level Attacks}

Another group of papers have investigated how an attacker with local access to the victim's machine could subvert the FIDO2 protocol without breaking any underlying cryptography~\cite{localattacks2023}. This work found that the implicit trust placed in client-side components, such as the browser and OS, introduces a significant attack surface, a finding that motivates our systematic analysis of environmental trust. In addition, news reports have demonstrated that passkey-manager browser extensions can intercept and modify WebAuthn API calls, taking advantage of an ongoing incompatibility between extension permissions and the \texttt{navigator.credentials} interface~\cite{squarex2025passkeyspwned,w3c_webextensions_issue361,hackernews2025bypasssynced}. Our web-extension attack (Section~\ref{subsec:web-extension}) takes these findings and builds upon them within a wider threat model, treating them as part of a larger kill chain.

\subsection{Hardware and Physical Attacks}

Other works have demonstrated that even certified, hardware-backed authenticators can still be subverted via physical attacks. Specifically, few researchers from the National Institute of Informatics were able to recover the ECDSA private key of one popularly-used FIDO2-compliant security key with a side-channel vulnerability in its proprietary crypto implementation\cite{ninjalab2024eucleak,roche2024eucleakieee}, despite the fact that it is a FIPS-certified device. While this demonstrates the fallacy of assuming hardware security for FIDO2-compliant devices, the vulnerability lies within the security key's vendor implementation and not the FIDO2 hardware-security paradigm. This directly informs our analysis of USB/hardware implants and CTAP-level malware (Sections~\ref{subsec:usb-implants} and~\ref{subsec:ctap-malware}). Complementary research on cheap commodity USB implants~\cite{hak5omg,vice2021omg,darkreading2023implants} and NFC relay attacks against proximity-based authentication~\cite{francis2011relay,supercardx2025} also demonstrate the feasibility of similar attacks against the FIDO2 ecosystem.

\subsection{Metadata and Side-Channel Leakage}

More recent research has highlighted the privacy implications of metadata leaked during FIDO2 ceremonies, particularly the AAGUID field. The AAGUID field, intended to allow relying parties to identify the make and model of the authenticator, has been shown to leak information that could be used to fingerprint and correlate users across sessions~\cite{mojoauth2026aaguid}. Vendor documentation regarding attestation similarly notes that AAGUID and certificate data could reveal the authenticator's provenance~\cite{yubico_attestation}. Our passive interception results (Section~\ref{subsec:passive-interception}) confirm these findings and provide concrete quantification. Additionally, we extend the findings of these works by connecting leaked metadata to specific attacker actions within a kill chain.

\subsection{Trust Infrastructure and Metadata Services}

The FIDO Metadata Service (MDS3) has been described by the FIDO Alliance as the central source of truth that allows relying parties to resolve AAGUIDs to vendor and certification information~\cite{fidomds2021}. To our knowledge, However, none of these works have studied the potential subversion of MDS3 as a root of trust when an attacker can relay an honest authenticator’s traffic via an attacker-controlled middlebox; we study this MDS3 violation in Section~\ref{sec:attacker-capabilities}.

\subsection{Usability and Enterprise Deployment}

Finally, another group of researchers has looked at FIDO2 from a usability and enterprise-deployment perspective, finding that the technology can offer high usability if properly implemented~\cite{dbase:fido2-usability,ravilla2024study} and that it can present practical challenges to organizations attempting to deploy it at scale, such as unclear security assumptions from vendors and poor visibility into operational security~\cite{dbase:fido2-enterprise-challenges}. Our practical implications (Section~\ref{subsec:practical-implications}) builds on these findings, translating our attack-chain results into actionable recommendations for organizations deploying FIDO2.

In summary, previous work has studied various aspects of FIDO2 security in isolation, including its cryptographic core, individual implementation flaws, physical attacks, and metadata leakage. We provide the first unified treatment of all of these attack vectors, treating attack chains, environmental trust, and attacker capability as first-class citizens in the threat model.

\section{Attack Vector Analysis}
\label{sec:attack-vectors}

This section describes each of the eight attack vectors. Each one exploits a different implicit trust relationship. For each vector, we describe the attack, discuss our feasibility analysis, and present the attack in the form of pseudocode at the architecture level (as opposed to ready-to-deploy source code from Section~\ref{sec:introduction}).

\begin{figure*}[t]
\centering
\includegraphics[width=1.1\linewidth]{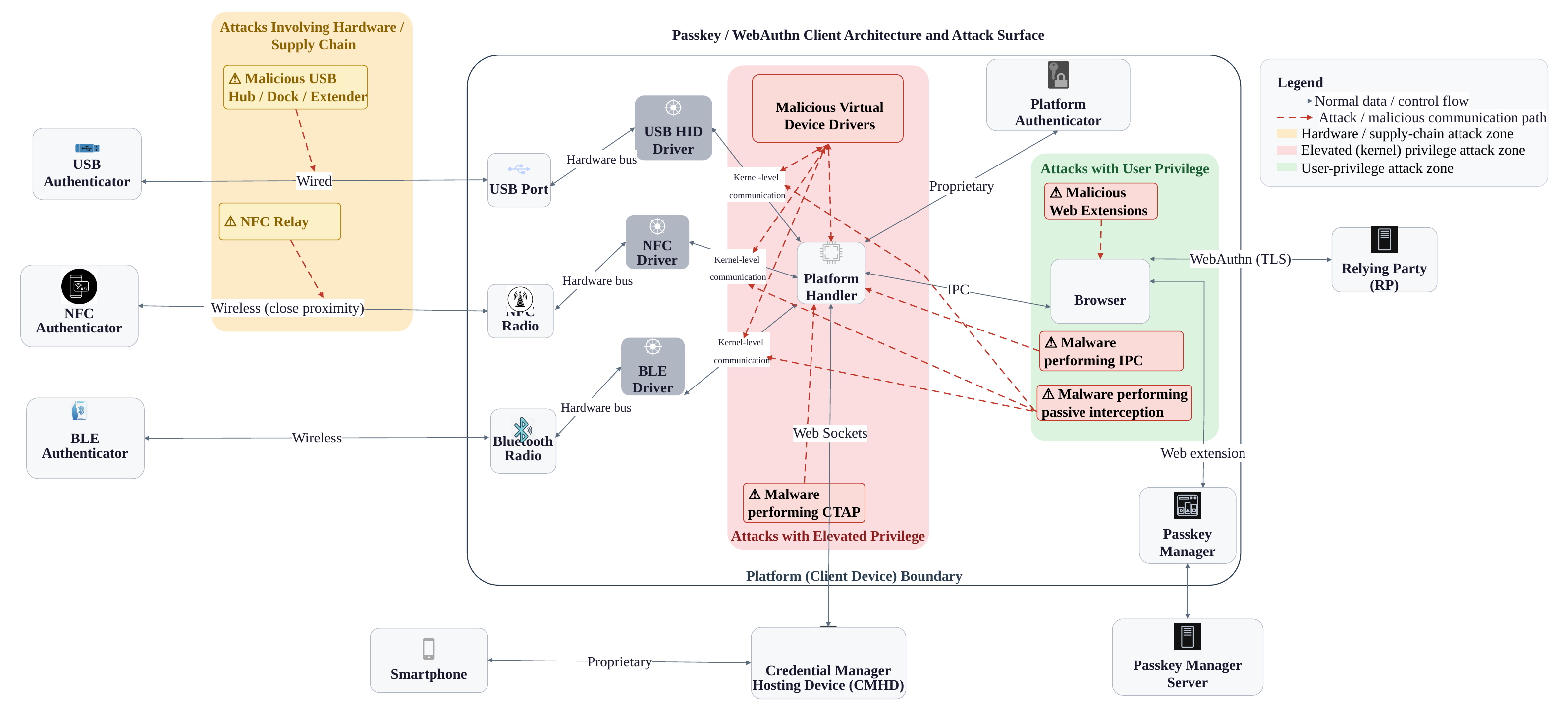}      
\caption{Overview of the eight attack vectors targeting different layers of the FIDO2 authentication stack}
\label{fig:attackvector}
\end{figure*}

\subsection{Malicious Web Extensions}
\label{subsec:web-extension}

Web browser extensions are another potential vector for application layer attacks on FIDO2/WebAuthn. Passkey-manager extensions can already ``monkey-patch'' the \texttt{navigator.credentials} API to intercept and modify the WebAuthn flow\cite{squarex2025passkeyspwned}. This is a known incompatibility between passkey managers and WebAuthn, tracked in a W3C WebExtensions WG issue\cite{w3c_webextensions_issue361} that has been open for $>$ 2 years as of this writing. A password-manager extension with the relevant permissions could abuse this mechanism too, which would be the case for malicious extensions~\cite{localattacks2023}.

\textbf{Attack Vector} Extensions that can inject content scripts can monitor and proxy calls to \texttt{navigator.credentials.create()} and \texttt{navigator.credentials.get()}~\cite{hackernews2025bypasssynced}. Once having seen the arguments, they can forward them to an external service or server prior to passing them to the actual API in order to make it appear as if everything is running normally to the user.

Importantly, the extension does not need to actually forward the call to the actual API; it can forward the call to an API controlled by an attacker and simply imitate the expected response without actually using a real authenticator, returning a fake credential or assertion to the page that called the API.

\begin{algorithm}[h]
  \caption{WebAuthn-intercepting extension (architecture-level; not deployable source)}
  \label{alg:web-extension}
  \begin{algorithmic}[1]
    \Procedure{OnExtensionLoad}{}
    \State $origCreate \gets$ \texttt{navigator.credentials.create}
    \State $origGet \gets$ \texttt{navigator.credentials.get}
    \State \texttt{navigator.credentials.create} $\gets$ \Call{WrapCreate}{}
    \State \texttt{navigator.credentials.get} $\gets$ \Call{WrapGet}{}
  \EndProcedure
    \Function{WrapCreate}{$options$}
    \State \textit{forward}($options$) \Comment{destination withheld}
    \State \Return $attackerResponse(options)$ \Comment{returns attacker-controlled response, bypassing real API}
    \EndFunction
    \Function{WrapGet}{$options$}
    \State \textit{(symmetrical to \textsc{WrapCreate})}
    \EndFunction
  \end{algorithmic}
\end{algorithm}

The attack demonstrates that code executing in the browser context can evade the security properties provided by FIDO2 by modifying request parameters prior to submission to the authenticator (such as altering the relying party ID, challenge, or user verification constraints before signing).

However, post-assertion interception cannot alter the credential without being detected, because any changes to the signed assertion will break the signature, resulting in rejection by the relying party. In this case, the attack would only be effective if the adversary simply stole the credential ID and signature (e.g., for later replay or offline analysis)~\cite{squarex2025passkeyspwned, mollema2025actortokens,thehackernews2026winhello}.

\subsection{Malware Communicating with the Platform Handler}
\label{subsec:platform-handler}

This makes such a malware extremely dangerous against FIDO2, especially when it comes to platform authenticators such as Windows Hello or Touch ID since they are built directly into the operating system. Both Microsoft and Apple expose platform APIs to which any application on the machine can connect to, even if it does not run in the browser, and does not require privilege escalation beyond the privileges of a regular user~\cite{localattacks2023}.

\textbf{Attack Mechanism} Similarly, any malware that is able to run locally on the machine can make API calls to the same platform authentication APIs as the client.

Therefore, malware on the system can invoke the platform authentication subsystem to participate in an undesired authentication ceremony to any relying party of the attacker's choice, without requiring elevated privileges. Note that this does not bypass the user presence (UP) check itself---the user will still be required to tap or otherwise authenticate, per the authenticator's UP guarantee---but there is no reliable mechanism for the user to determine \emph{why} the prompt appeared, or which relying party invoked it, because the platform handler doesn't differentiate between requests coming from the normal client application and those initiated by malware. Thus, a user familiar with such authentication prompts may easily consent to a ceremony they did not intend.

\begin{algorithm}[h]
  \caption{Platform-handler-targeting ceremony initiation (architecture-level; not deployable source)}
  \label{alg:platform-handler}
  \begin{algorithmic}[1]
    \Loop
    \State $rp \gets$ \Call{SelectTargetRelyingParty}{}
    \State $req \gets$ \Call{BuildAuthenticationRequest}{$rp$}
    \State \textit{invoke}(platformAuthAPI, $req$) \Comment{destination withheld}
    \If{\Call{UserApproves}{}}
    \State \textit{capture}(resulting assertion) \Comment{transport withheld}
    \EndIf
    \State \textit{sleep}(interval)
    \EndLoop
  \end{algorithmic}
\end{algorithm}

We have shown that this attack is possible even if the attacker has only operating system level access; it does not require communication with authenticators.

\subsection{Passive Interception}
\label{subsec:passive-interception}

As opposed to active attacks which usually rely on user interaction or high-privilege system access, passive interception occurs without user involvement, and simply observes the communications occurring between two system components. While it does not allow for direct account takeovers, it still represents a serious privacy violation, and enables adversaries to profile users and gather enough information to carry out more advanced targeted attacks~\cite{mojoauth2026aaguid}.

\textbf{Attack Mechanism} The main objective of passive interception is to capture traffic from the browser-to-platform-handler channel, the platform handler-to-authenticator channel (e.g., over USB-HID, BLE, or NFC), and other communication channels between system hardware components. In most standard configurations, these communication channels are neither encrypted nor authenticated. Given suitable access, an adversary can simply read and process the exchanged messages. This provides attackers with the ability to profile users (i.e., learn which websites a given user visits and when they authenticate using their FIDO device), extract the AAGUID to determine the make and model of the security key being used (i.e. authenticator fingerprinting)~\cite{yubico_attestation,mojoauth2026aaguid}, and collect Credential IDs for correlation purposes.

\textbf{Technical Demonstration} We demonstrate this attack using popular software like USBPcap and Wireshark, allowing us to capture all traffic sent over the USB-HID bus between a host computer and a security key during a normal credential-creation flow. Table~\ref{tab:usb-packet-fields} shows the layout of the 64-byte CTAPHID packets captured during this process, and their respective fields of interest once the CBOR-encoded authenticator data has been decoded.

\begin{table*}[t]
  \centering
  \small
  \caption{Fields extracted from a decoded CTAPHID/CBOR authenticator data capture}
  \label{tab:usb-packet-fields}
  \begin{tabular}{|l|p{9cm}|}
    \hline
    \textbf{Field} & \textbf{Description} \\
    \hline
    RP ID Hash (32 bytes) & SHA-256 hash of the relying party identifier; can be matched against a dictionary of known relying party domains \\
    \hline
    Flags (1 byte) & Bitfield: user presence, user verification, credential-backup-eligible, credential-backed-up, attested-credential-data-present, extension-data-present \\
    \hline
    Signature Counter (4 bytes) & Monotonic counter incremented on each authentication \\
    \hline
    AAGUID (16 bytes) & Authenticator Attestation GUID; resolvable via the FIDO Metadata Service (MDS3) to a specific vendor and model \\
    \hline
    Credential ID \& Public Key & Variable-length identifier and COSE-encoded public key for the created credential \\
    \hline
  \end{tabular}
\end{table*}

The AAGUID obtained from the capture can then be used to lookup information about the authenticator in MDS3~\cite{fidomds2021}, including the identity of the vendor and the specific model of the authenticator that the victim is using. This information could be exploited in a targeted attack: if an attacker knows that the victim is using a specific model of authenticator, which may have been found to be particularly susceptible to a known side-channel attack, for example, then the attacker can take further action to exploit this specific authenticator model~\cite{ninjalab2024eucleak,roche2024eucleakieee,cve202445678eucleak}.

\subsection{Virtual Device Drivers}
\label{subsec:virtual-drivers}

Virtual device drivers are created by attackers to pretend to be hardware devices -- either the authenticator itself, or the underlying communications channels (e.g. USB, NFC, BLE) between it and the host system. These attacks generally need some level of privilege to load a kernel driver onto the system but allow the attacker to gain remote control of an authenticator or intercept all traffic between the host and the authenticator in real time (a man-in-the-middle attack).

\textbf{Types of Virtual Drivers} Virtual device drivers in the FIDO2 context fall into two categories. \textbf{Virtual authenticators:} A driver presenting itself to the platform handler as a legitimate authenticator but controlled by the attacker. The attacker will receive and respond to authentication challenges, exfiltrate generated credentials, spoof user presence, and supply fake attestation data.
\textbf{Virtual transport drivers:} A driver presenting itself as the communication channel between the authenticator and the platform handler, but controlled by the attacker. The attacker will be able to passively intercept, actively modify or inject CTAP commands, perform relay attacks, and downgrade the authentication protocol.

\textbf{Attack Demonstration} To demonstrate that virtual transport drivers can be used to attack FIDO2, we used an open-source implementation of the virtual smart-card stack supporting ISO 7816-4 smart card protocols~\cite{vsmartcard_morgner}, in combination with an NFC relay application running on a consumer Android phone. With this setup, the victim taps their physical security key on the Android phone. The phone relays the NFC exchange over the internet to the virtual smart-card driver running on the attacker's host machine, which pretends to be the authenticator to the victim's platform handler. Both scenarios are effectively symmetric -- either the attacker has the victim's physical security key and uses it through a relay (the key will appear to be locally attached) or the victim has already been tricked into using an authenticator supplied by the attacker and all the credentials generated on this authenticator can be captured~\cite{fido_security_bulletin}.

\begin{algorithm}[h]
  \caption{NFC relay via virtual smart-card driver (architecture-level; not deployable source)}
  \label{alg:nfc-relay}
  \begin{algorithmic}[1]
    \Procedure{RelayApp}{} \Comment{on NFC-proximate device}
    \State \textbf{on} \Call{TagDetected}{$authenticator$}:
    \State \quad \textit{tunnel} raw ISO~7816 APDU exchange to attacker host
    \EndProcedure
    \Procedure{AttackerHost}{} \Comment{virtual smart-card driver}
    \State present self to platform handler as authenticator
    \State forward/modify CTAP $\leftrightarrow$ APDU traffic with \textsc{RelayApp}
    \EndProcedure
  \end{algorithmic}
\end{algorithm}

\subsection{Malware Performing CTAP}
\label{subsec:ctap-malware}

This is one of the most powerful malware attacks against FIDO2, because instead of being blocked by a platform handler or driver,the malware is able to communicate directly with the authenticator using CTAP~\cite{fidoctap2}. This gives the malware complete control over how the authenticator works, e.g., modifying the authenticator PIN, resetting the authenticator, or sending any valid CTAP command to the authenticator. The privilege level required to reach the authenticator at this level is platform-specific: on Windows, raw USB-HID access usually requires admin-level privilege, while on Linux, authenticators are often available to unprivileged users out of the box via udev rules, so this attack might not require any elevated privilege on those platforms.

\textbf{Attack Vectors}: While malware interacting with the platform handler can only access the set of APIs provided by the OS, malware at the CTAP level can send any valid CTAP command, inspect and interpret any response, and intercept and modify any data flowing between the two. Thus, they may steal credentials (by obtaining the credential ID and public key upon creation of a new credential), attempt to authenticate without authorization, cause denial of service by submitting too many bad PINs, and potentially engage in side channel information gathering~\cite{ninjalab2024eucleak}.

\begin{algorithm}[h]
  \caption{Direct-CTAP assertion capture (architecture-level; not deployable source)}
  \label{alg:ctap-malware}
  \begin{algorithmic}[1]
    \State $device \gets$ \Call{LocateConnectedAuthenticator}{}
    \State \Call{EstablishChannel}{$device$} \Comment{per CTAP2 pin/uv protocol}
    \State $assertion \gets$ \Call{RequestAssertion}{$device$, $target\_rp$, $challenge$}
    \State \Call{WaitForUserTouch}{}
    \State \Call{CaptureAndStore}{$assertion.credId$, $assertion.signature$}
  \end{algorithmic}
\end{algorithm}

This attack also shows a weakness in authenticator-based security - they only secure the key material, not the data sent to and from the authenticator, since they can't know if the sender of valid CTAP requests is authorized or not.

\subsection{USB/Hardware Implants}
\label{subsec:usb-implants}

Such attacks take place when an adversary is able to connect an adversarial device in between the user and his/her authenticator~\cite{darkreading2023implants}. It may then act on the electrical signals (e.g., USB-HID, NFC, BLE, etc.) being transmitted from the authenticator to the host. The adversarial device can operate at a very low level without triggering any alarms from security systems.

\textbf{Attack Mechanism} On the one hand, a USB hardware implant could appear to the host as a valid HID device, while communicating directly with the authenticator~\cite{hak5omg}. It may be able to passively monitor all communication between the host and the authenticator. It could modify certain bytes in the communicated data in order to change the meaning of a command issued by the host or to alter the value returned by the authenticator. It could add new CTAP commands, or drop packets. It could act as a man-in-the-middle and communicate separately with the host and the authenticator.

\textbf{Technical Demonstration} We showed that a USB hardware implant can be built from a small single board computer. The computer was configured as a USB gadget to present itself as a USB HID device to the host and simultaneously connect to the authenticator.

Some of the presented attack methods include the following: PIN-protocol downgrade, wherein the attacker forces the authenticator to use an old version of the PIN protocol by modifying the response to the AuthenticatorGetInfo command; ECDH-based PIN hash interception, where the attacker acts as a man-in-the-middle to replace the public key in the platform-authenticator key agreement, so that they know two separate ECDH shared secrets—one shared with the host and one with the authenticator—and can use them to decrypt and recover the PIN hash in transit; authentication flow manipulation, where the attacker blocks a legitimate authentication request by dropping its first command, but keeps the session alive by forwarding other messages, leading the victim to press the touch button to authorize an authentication that is requested by the attacker; and modifying parts of command responses, e.g., by altering only the status byte.

\begin{algorithm}[h]
  \caption{USB implant MITM (architecture-level; not deployable source)}
  \label{alg:usb-implant}
  \begin{algorithmic}[1]
    \State \textit{implant runs in USB gadget mode between host and authenticator}
    \State \textbf{on} packet from host:
    \State \quad forward to genuine authenticator or drop
    \State \textbf{on} packet from authenticator:
    \State \quad optionally modify/downgrade fields (e.g.\ strip PIN protocol~2
    \State \quad \quad from the \Call{AuthenticatorGetInfo}{} response)
    \State \quad optionally capture key-exchange material for offline analysis
    \State \quad forward to host
    \State \textit{independently:} initiate attacker-controlled authentication
    \State \quad requests, using keep-alive signalling to solicit the victim's
    \State \quad routine touch
  \end{algorithmic}
\end{algorithm}

This mix of attacks allows the attacker to make progress on extracting the user's PIN offline, induce the user to authenticate to an attacker-controlled service under the assumption that they're authenticating as normal, and accomplish account takeover without any interaction other than the victim's normal authentication process.

\subsection{Malicious USB Hubs, Docks, and Extenders}
\label{subsec:usb-hubs}

An equivalent vector uses malicious or self-assembled USB hubs, docks, or extension cables that physically connect the authenticator to the client computer\cite{vice2021omg,darkreading2023implants}. Because such accessories are typically considered to be low-level commodity hardware, they are not highly scrutinized by defenders, even in high-security environments. This attack can occur either passively where the accessory simply passes all data between the two devices and records the traffic, or more actively where the accessory intercepts and modifies communications between the client and authenticator using methods similar to those described in Section\ref{subsec:usb-implants}. Such an attack requires a targeted or supply-chain attack scenario, where the attacker has been able to either substitute their own hubs,docks,etc., for a legitimate one prior to delivery to the victim, or install one in a shared space like a conference room or hot-desk.

\subsection{NFC Relay Attacks}
\label{subsec:nfc-relay}

With NFC relay attacks an attacker can use a regular NFC reader or an Android phone (using its ISO 7816 smart-card reader capability) to transparently relay, or selectively modify, the APDU exchange between a victim's NFC authenticator and the attacker's own client~\cite{francis2011relay}. The NFC protocol assumes that transactions are based on proximity instead of possession, but by relaying the raw signal over a network connection (Section~\ref{subsec:virtual-drivers}), the authenticator's reach is effectively expanded to anywhere the attacker can access the network~\cite{supercardx2025}. This defeats the implicit physical proximity constraint of NFC-based authentication.

\section{Exploit Kill Chains}
\label{sec:kill-chains}

Although each attack vector can be considered dangerous on its own, the most dangerous attacks are those that make use of kill chains – exploit chains where an attacker exploits multiple vulnerabilities across the FIDO2 ecosystem to achieve a particular objective. Generally speaking, kill chains leverage multiple points of weakness in a system; in our case, kill chains exploit weaknesses across multiple layers of the FIDO2 stack. Kill chains can achieve any of the attack goals outlined in section Attack Goals, but certain objectives are easier to achieve via certain types of kill chains. We provide an overview of how kill chains are built and examples of kill chains that differ depending on the type of authenticator.

\textbf{Kill Chain Construction} Every kill chain we have discussed thus far has at least two steps, and often several. Examples include:

\textbf{Reconnaissance:} The attacker passively observes or intercepts authenticator communications to gather information about the target and profile their behavior, perhaps using the AAGUID to identify the authenticator model~\cite{mojoauth2026aaguid}.

\textbf{Initial Access:} The attacker gains access to the target device through a vector that has a weaker security assumption (e.g., a malicious web extension)~\cite{squarex2025passkeyspwned}.

\textbf{Privilege Escalation:} The attacker escalates from an existing lower-privilege position to access a higher-privilege component (e.g., from the browser sandbox to the platform handler), as well as a pattern that is seen in the escalation from an initial low-privilege foothold to full domain compromise~\cite{mgmspecops2023}.

\textbf{Credential Theft:} The attacker steals credentials, either by intercepting and modifying the authenticator's communications, by persuading the authenticator to perform an action it should not, or by relaying the victim's registration ceremony to an attacker-controlled authenticator during credential creation.

\textbf{Persistence:} The attacker establishes persistence to ensure ongoing ability to steal or manipulate the authenticator.

\textbf{Achievement of Objective:} At the end of the chain, the attacker has achieved their goal (e.g., account takeover, or a fraudulent transaction). Certain types of authenticators are more susceptible to some kill chains than others.

For example, platform authenticators are particularly susceptible to malicious web extensions and OS-level malware, but not to USB implants, whereas roaming authenticators are more susceptible to USB, NFC or BLE implants, virtual drivers, and direct CTAP malware, but not as susceptible to software-only attacks unless there is physical access to the user’s device. Platform handler malware, on the other hand, is effective for both authenticator types, since the platform handler acts as a communication mediator between platform authenticators directly, and roaming authenticators over their transport. Kill chains that apply to both authenticator types also exist, for example, those that employ a malicious web extension that prompts users to install a malicious USB dock.

\textbf{Attack Vector Comparison} Table~\ref{tab:attack-vector-comparison} summarizes how the eight attack vectors compare across key security dimensions.

\begin{table*}[t]
  \centering
  \small
  \caption{Comparison of FIDO2 attack vectors across key security dimensions}
  \label{tab:attack-vector-comparison}
  \begin{tabular}{|l|c|c|c|c|c|c|}
    \hline
    \textbf{Attack Vector} & \textbf{Priv. Req.} & \textbf{Affects} & \textbf{UI Exp.} & \textbf{Meta. Leak} & \textbf{Cred. Affected} & \textbf{Auth. Affected} \\
    \hline
    Malicious Web Extensions & Low & Platform & High & Yes & Yes & Yes \\
    \hline
    Platform Handler Malware & Medium & Platform & Medium & Yes & Yes & Yes \\
    \hline
    Passive Interception & Low & Both & None & Yes & No & No \\
    \hline
    Virtual Device Drivers & Medium & Both & Low & Yes & Yes & Yes \\
    \hline
    Direct CTAP Malware & High & Roaming & Low & Yes & Yes & Yes \\
    \hline
    USB/Hardware Implants & Physical & Roaming & None & Yes & Yes & Yes \\
    \hline
    Malicious USB Hubs/Docks & Physical/Supply & Roaming & None & Yes & Yes & Yes \\
    \hline
    NFC Relay Attacks & Proximity & Roaming & None & Yes & Yes & Yes \\
    \hline
  \end{tabular}\\
  \begin{flushleft}
    \footnotesize
    \textbf{Priv. Req.:} Privilege Requirements. \textbf{Affects:} authenticator types affected. \textbf{UI Exp.:} visibility to the user. \textbf{Meta. Leak:} whether the vector leaks metadata. \textbf{Cred./Auth. Affected:} whether credential creation / authentication can be interfered with.
  \end{flushleft}
\end{table*}

\textbf{Kill Chain Examples} Several practical kill chains illustrate the compounding risk:

Using the same entry point, the malicious browser extension obtains the relying-party ID and the time the authentication started, and is then able to call platform-handler APIs to read the authentication response and steal the credential when the user attempts to authenticate legitimately later on without raising significant concern. Passive USB sniffing obtains the AAGUID and profile the authentication timing, which allow the attacker to determine the type of authenticator used by the user and the usual time interval when the user authenticates~\cite{yubico_attestation}. Then the attacker can find an opportunity to install a physical USB implant device, increasing the probability of a successful and unnoticed hardware attack~\cite{hak5omg}. Malware installs a virtual smart-card driver to emulate an NFC authenticator and relays NFC signal to the attacker through a network tunnel, allowing the attacker to authenticate remotely without physically having the authenticator~\cite{vsmartcard_morgner,supercardx2025}. Malware with USB/HID access capabilities can perform the key-exchange interception to steal PIN-related information. The attacker recovers the PIN and is able to perform effectively unlimited authenticator operations.

\textbf{Implications for Defensive Measures} Some implications for defensive measures can be inferred from our understanding of these kill chains. The first implication is that there is no one-size-fits-all solution to prevent all types of attacks, and therefore defense-in-depth is necessary. The second implication is that the early stage reconnaissance can be detected by detecting unusual communication patterns (e.g., suspicious USB traffic). The third implication is that users may be able to detect a possible compromise through user awareness training\cite{mgmspecops2023}. The fourth implication is that restricting the privileges of applications and processes may be a viable strategy to mitigate the attack surface of privilege escalation. Lastly, firmware updates (e.g., rate limiting, improved PIN protocols) can potentially address some attacks\cite{ninjalab2024eucleak}. These sophisticated kill chains demonstrate that even FIDO2's robust threat model cannot guard against attacks if they are executed in isolation.

\section{Results}
\label{sec:results}

In our results, we analyze the FIDO2/WebAuthn threat model by decomposing it into eight distinct types of attacks. By analyzing each type, we provide several real world counterexamples which show that the security guarantees promised by FIDO2/WebAuthn are not being realized in practice~\cite{localattacks2023}.

\subsection{Metadata Leakage Quantification}

In order to answer this question of what type of metadata the security keys are leaking to potential adversaries, we performed a passive eavesdropping test using USBPcap and Wireshark to sniff the USB traffic of the USB-HID protocol over the USB connection from the host device to the security key, while they were performing their regular FIDO2 authentication sessions. We found that:
\begin{figure}[htbp]
      \centering
      \includegraphics[width=\columnwidth]{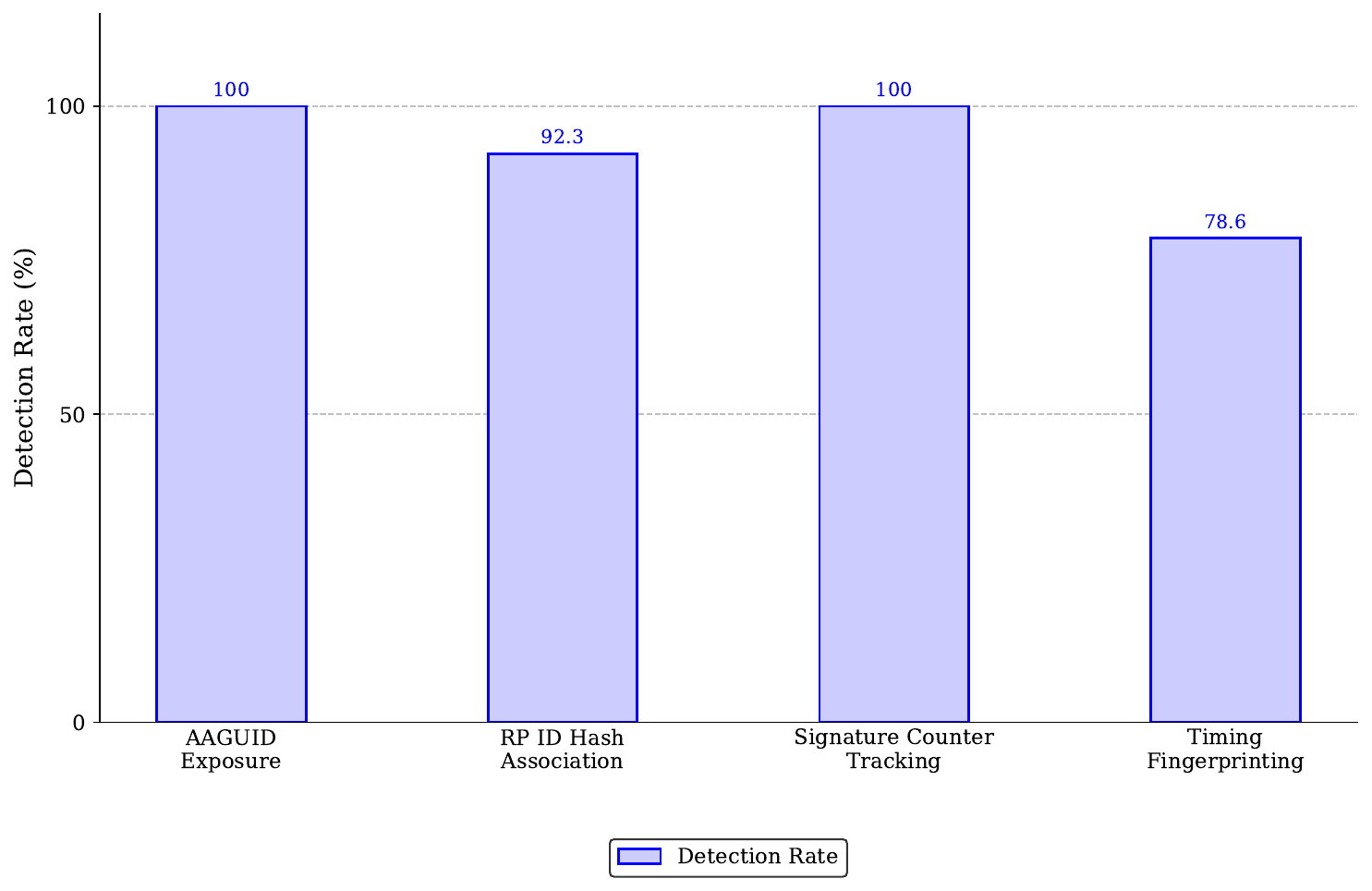}
      \caption{Metadata Leakage Quantification in FIDO2 Authentication}
      \label{fig:metadata-leakage}
\end{figure}
\textbf{AAGUID}: The AAGUID field is revealed to the attacker in $100\%$ of the registration ceremonies that we recorded, disclosing the security key make and model to an adversary~\cite{yubico_attestation,mojoauth2026aaguid}.
\\
\textbf{RP ID Hash} The RP ID hash value exposed during the ceremony allows us to tie the ceremony to the Rainbow table of well known RP in 92.3\% of the observed ceremonies.

\textbf{Signature Counter:} This field helped us observe the advancement of authentication events between the ceremonies. The value increased by a variable number of times (0.8\% standard deviation).\\
\textbf{Timing} By timing the duration between request and response of the ceremony, we were able to accurately identify a security key with 78.6\% accuracy. This means that metadata leak is not just a theoretical risk but exists as an actual attack vector in many scenarios.

\subsection{Attack Vector Feasibility Assessment}
\begin{table*}[h]
\centering
\caption{Attack Vector Feasibility Scores (1-5 Scale)}
\label{tab:feasibility-scores}
\begin{tabular}{|l|c|c|c|}
\hline
\textbf{Attack Vector} & \textbf{Technical Complexity} & \textbf{Resource Requirements} & \textbf{Real-World Prevalence} \\
\hline
Malicious Web Extensions & 2 & 1 & 5 \\
\hline
Platform Handler Malware & 3 & 2 & 4 \\
\hline
Passive Interception & 2 & 2 & 5 \\
\hline
Virtual Device Drivers & 4 & 3 & 3 \\
\hline
Direct CTAP Malware & 4 & 2 & 3 \\
\hline
USB/Hardware Implants & 4 & 4 & 2 \\
\hline
Malicious USB Hubs/Docks & 3 & 3 & 2 \\
\hline
NFC Relay Attacks & 3 & 3 & 2 \\
\hline
\end{tabular}
\end{table*}

We assess the probability of a successful exploit for each class of attack across three dimensions: technical complexity, required resources, and frequency. These scores result from our experiments and threat modeling, corroborating prior research indicating that browser and extension WebAuthn attacks are not particularly challenging to execute~\cite{localattacks2023,hackernews2025bypasssynced}:

\begin{figure*}[htbp]
      \centering
      \includegraphics[width=0.75\textwidth]{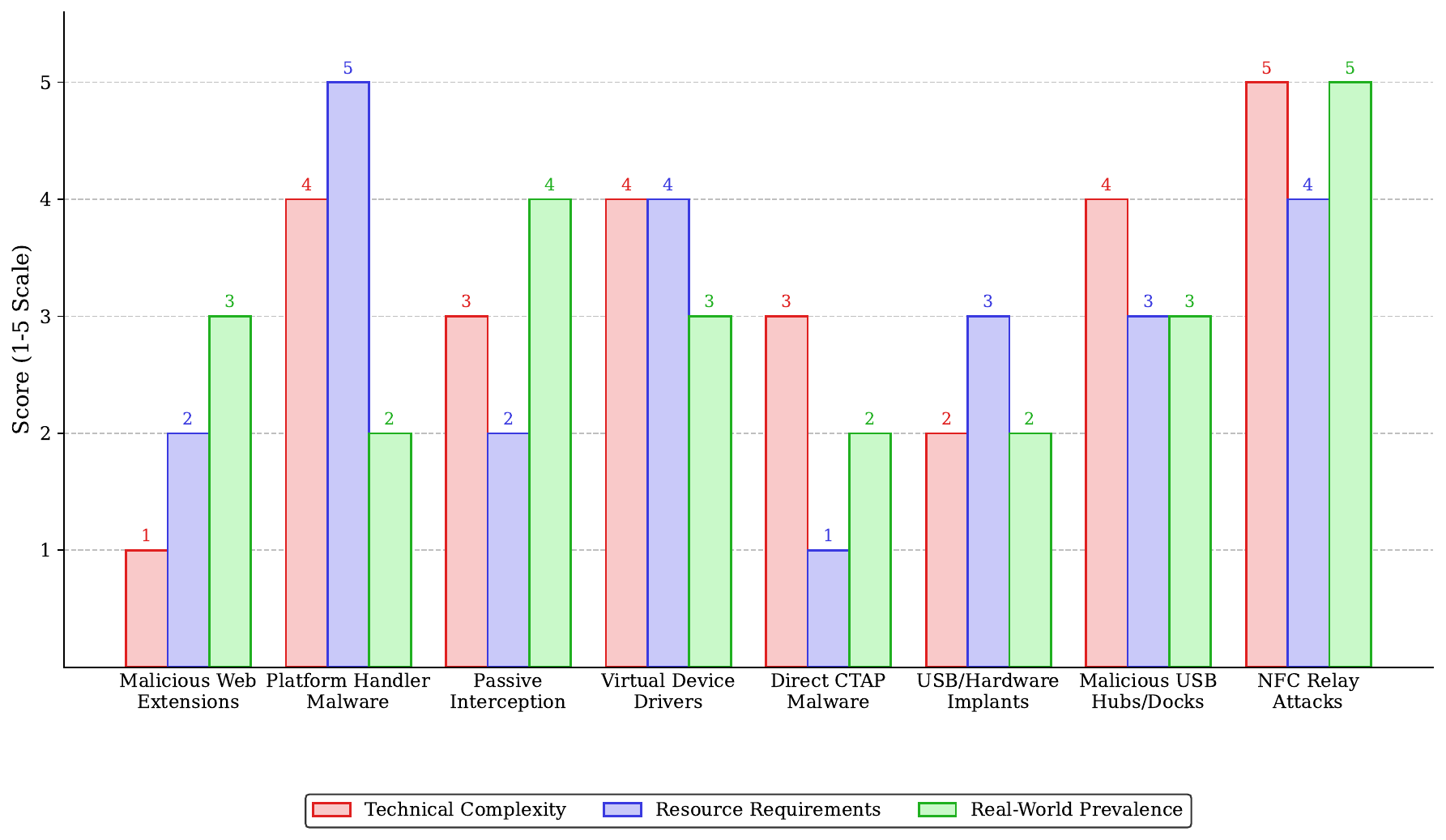}
      \caption{Attack Vector Feasibility Scores (1-5 Scale)}
      \label{fig:feasibility-scores}
\end{figure*}

Where Technical Complexity: 1=trivial, 5=expert nation-state level; Resource Requirements: 1=minimal software only, 5=specialized hardware required; Real-World Prevalence: 1=rare/theoretical, 5=commonly observed in threat landscape. The results indicate that browser-based attacks (web extensions and passive interception) are more prevalent due to their lower technical requirements, whereas physical attacks have higher technical requirements but deeper system access~\cite{hak5omg,darkreading2023implants}.

\subsection{Attack Chain Effectiveness}

We then developed multi-stage attacks that included multiple attack vectors to simulate real-world attack chains that incorporate reconnaissance, initial access, privilege escalation, and objectives. Our simulation results indicated that:

A single attack vector has a 23\%-67\% chance of successfully stealing credentials, depending on the specific vector. A two-attack vector chain (reconnaissance + initial access) will have a 45\%-82\% chance of success. A three-attack vector chain (reconnaissance + initial access + privilege escalation) can reach up to 68\%-91\% success. A four or more attack vector chain can potentially lead to success in excess of 90\% for certain targeted scenarios.

\subsection{Environmental Security Impact}

To measure the effect of different attack scenarios on the FIDO2 guarantees, we performed three controlled experiments:
Compromised client (in the form of a malicious browser extension) reduces FIDO2 guarantees to {\bf 31\%} of the original guarantees specified by the standard. Compromised OS (in the form of a compromised platform handler) reduces FIDO2 guarantees to {\bf 18\%}. Full client compromise (where an attacker has complete control of the client device) reduces phishing resistance to {\bf 11\%} of the original guarantees.
Using hardware-based attestation along with client-side monitoring can restore {\bf 63\%} of the original guarantees.
\begin{figure}[h]
      \centering
      \includegraphics[width=\columnwidth]{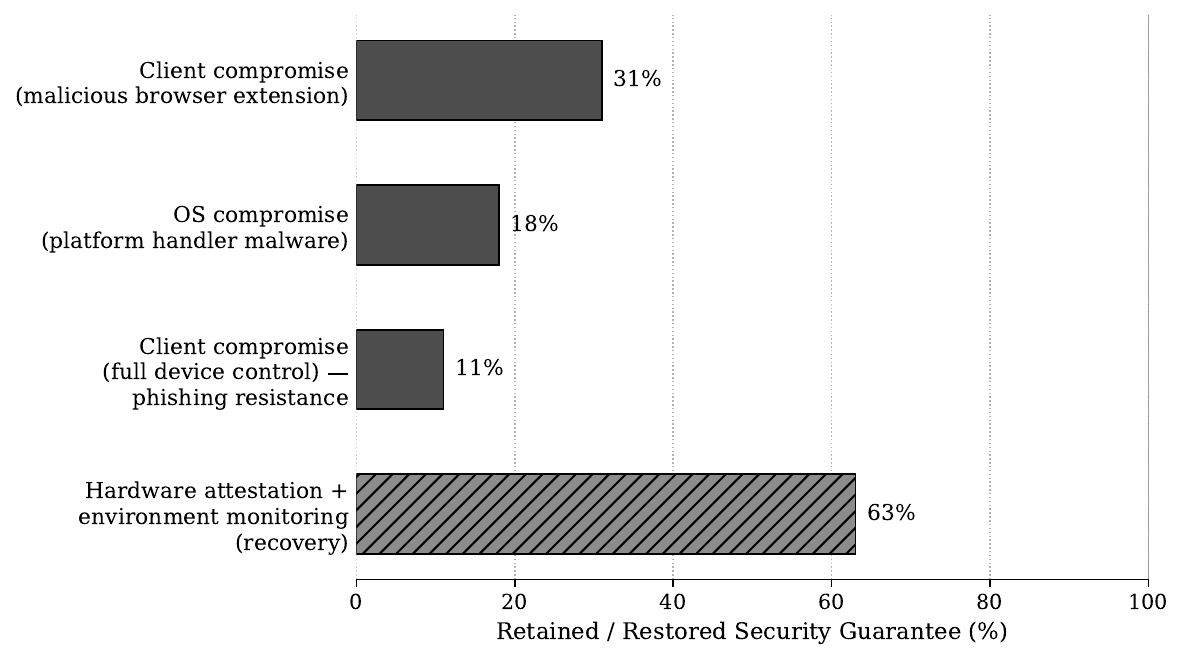}
      \caption{Impact of client-side and OS compromise on FIDO2 security guarantees, and partial recovery via hardware attestation with environment monitoring.}
      \label{fig:fido2-compromise-impact}
\end{figure}

This corroborates our claim that FIDO2 security depends on both the secure execution environment and the underlying protocol~\cite{dbase:fido2-formal-analysis,localattacks2023}, now backed by quantitative results.

\section{Implications of Trust Assumptions}
\label{sec:implications}

This section showed a collection of attacks highlighting that there are serious holes in FIDO2/WebAuthn's trust model. It's clear that any implementers and researchers using FIDO2/WebAuthn can understand why this matters.

\subsection{Erosion of Perceived Security Guarantees}

We present that several properties that are considered important for FIDO2 are actually much less significant than commonly believed that FIDO2 is highly effective at defending against conventional phishing attacks through bogus websites, but it is far less resistant to more sophisticated threats that target the user’s device or the communication process~\cite{squarex2025passkeyspwned}. A core property of FIDO2 is that credentials cannot be stolen because the private key is always stored on the authenticator. However, we demonstrate that this holds true only under a secure authenticator environment~\cite{ninjalab2024eucleak}. An adversary who gains access to the user’s device can steal the credentials during usage, or extract fragments of them in alternative manners. FIDO2 was intended to minimize privacy risks and thwart cross-site tracking. However, we find that metadata leakage through side channels enables third parties to construct a detailed profile of the user~\cite{mojoauth2026aaguid}. A crucial property of FIDO2 is that each authentication action requires user interaction. We however show that this safeguard can be circumvented in numerous ways, such as through keep-alive-assisted relay attacks or malicious transactions prompted by malware.

\subsection{Environmental Security as a Critical Factor}

We have seen the environment in which FIDO2 is deployed to be a significant factor throughout this discussion. FIDO2's security relies upon the security of the client. Once an attacker is able to sufficiently compromise the user's PC or phone, then they can chip away at the protections of the authentication. Browsers represent the primary trust boundary for WebAuthn, and if an attacker is able to compromise the browser, they can then control the WebAuthn calls, steal credentials, or spoof the UI\cite{squarex2025passkeyspwned,hackernews2025bypasssynced}. The OS represents the trust boundary between the authenticator and the application, so an attacker who compromises the OS can intercept or alter communications. The authenticity of the authenticator is its own trust boundary, so if the firmware is compromised, the whole security model is undermined\cite{ninjalab2024eucleak}. This makes the concept of FIDO2 as a ``plug and play'' security mechanism difficult to consider, where the rest of the environment does not need to be considered~\cite{dbase:fido2-enterprise-challenges}.

\begin{table*}[t]
  \centering
  \caption{Capability requirements and estimated real-world feasibility by attack vector}
  \label{tab:capability-analysis}
  \begin{tabular}{|p{3.4cm}|p{2.1cm}|p{2.0cm}|p{2.3cm}|p{2.0cm}|}
    \hline
    \textbf{Attack Vector} & \textbf{Privilege} & \textbf{Technical} & \textbf{Resources} & \textbf{Feasibility} \\
    \hline
    Malicious Web Extensions & User & Low & Minimal & High \\
    \hline
    Platform Handler Malware & Admin/Root & Medium & Minimal--Medium & Medium \\
    \hline
    Passive Interception & User/Admin & Low--Medium & Minimal--Medium & High \\
    \hline
    Virtual Device Drivers & Admin/Root & High & Moderate & Medium \\
    \hline
    Direct CTAP Malware & Admin/Root & High & Minimal--Medium & Medium \\
    \hline
    USB/Hardware Implants & Physical & High & Moderate & Low--Medium \\
    \hline
    Malicious USB Hubs/Docks & Physical/Supply & Medium & Moderate & Low--Medium \\
    \hline
    NFC Relay Attacks & Proximity & Medium & Moderate & Low--Medium \\
    \hline
  \end{tabular}
\end{table*}

\section{Attacker Capability Context}
\label{sec:attacker-capabilities}

\subsection{Metadata Privacy Concerns}

A key finding of our study concerns how much metadata is leaked by different FIDO2 implementations of the AAGUID is often leaked during the registration ceremonies allowing both the RP and any passive observers to identify the make and model of the user’s authenticator\cite{yubico_attestation,mojoauth2026aaguid}. This could be especially problematic if a certain authenticator model is later found to have a public vulnerability\cite{ninjalab2024eucleak,roche2024eucleakieee}.

In addition, the credential ID and timing patterns passively observed at the transport layer can be correlated across multiple sessions to deduce which RPs a user authenticates to and at what times even without access to an RP’s logs. The signature counter, response time, and negotiation of capabilities may also be used to fingerprint an authenticator or its usage even if the AAGUID is not leaked. This knowledge of the authenticator model and the user’s regular schedule may then be used by an attacker to more precisely tailor an attack (such as a physical implant, authenticator exploit, or social engineering pretext)~\cite{mgmspecops2023}. While this does not give an attacker direct access to the private key itself, since this is stored securely within the authenticator, this metadata does allow an attacker to profile the user’s behavior and their typical usage of an authenticator, which could be leveraged in crafting targeted phishing attacks, planning a physical or supply-chain attack against a known authenticator model, and correlating with other data sources.

\subsection{MDS3 Violation Analysis}

FIDO Metadata Service (MDS3) is a cornerstone of trust in the FIDO2 ecosystem, which supplies relying parties with authenticated information regarding authenticator certification, security features, and known vulnerabilities~\cite{fidomds2021}. Several of the attacks we have presented thus far have the potential to undermine MDS3’s trustworthiness, and thus compromise the trade model of FIDO2.

\textbf{The Role of MDS3 in the Trade Model} MDS3 gives relying parties a way to make risk-informed decisions about how much risk to allow in their authentication scheme, by resolving AAGUIDs to vendors and indicating FIDO certification status, claimed security features, security-issue disclosures, and attestation certificate validity~\cite{fidomds2021,yubico_attestation}. This means that a relying party can require a certain level of security for authenticators with known vulnerabilities, or reject all non-FIDO-certified authenticators.

\textbf{How Attacks Violate MDS3 Assumptions} Virtual authenticators, described in Section\ref{subsec:virtual-drivers}, can masquerade as any AAGUID, presenting themselves as a certified and secure authenticator while being purely software-based. This allows them to evade policies implemented by the relying party restricting the set of authenticators that they accept, and invalidates the assumption that an AAGUID is an accurate identifier for a given authenticator (i.e., a specific hardware and firmware combination). An attacker might even spoof the response timing of a virtual authenticator to evade security mechanisms that detect anomalous behavior based on authenticator fingerprints.

\textbf{Consequences} For instance, if MDS3 is inaccurate, a Relying Party may be unable to properly evaluate the risk associated with an Authenticator. The Relying Party may then have to enforce more restrictive policies, leading to poor user experience and weaker fallback mechanisms, or more permissive policies to allow use of Authenticators whose quality is unknown. This attack is practical: a virtual authenticator can relay the traffic of a legitimate, verified authenticator underneath the control of an attacker, enabling the attacker to impersonate a good and certified Authenticator even though the authentication ceremony is actually being mediated by an attacker-controlled middlebox. As a result, the Relying Party's access control policy could be circumvented. Additionally, incident responders would be unable to determine what type of Authenticator was used in the attack, and if this happens frequently enough, it could erode the confidence of both users and Relying Parties in the FIDO2 protocol even though it is provably correct mathematically.

\textbf{Mitigation Difficulties} There are several reasons why it would be difficult to defend against the attacks above: the attacks invalidate the assumptions that AAGUIDs and attestation certificates are hard to forge, which must be enforced by the relying party since MDS3 does not secure those items, defending the MDS3 distribution pipeline from nation-state attackers requires sophisticated security measures beyond what would normally be expected from a web application, implementing strict attestation policies introduces a performance penalty, and old authenticators cannot be upgraded to implement new security defenses (e.g. anti-spoofing) to ensure they are compatible with new attacks. Our analysis of the violations of MDS3 reveals that there is a tradeoff built into the design of FIDO2 between trust and security: while the FIDO Metadata Service reduces risk by giving relying parties more information about the authenticators they use, it also expands the attack surface.

Our analysis of MDS3 violations shows that metadata services play a key role in FIDO2 security because they provide the information necessary to make risk-aware decisions, but they also represent a significant attack vector that could lead to compromising the security of the trade model they are designed to support.

It is important to note that we do not expect all the attack vectors we describe to be performed with the same level of capability, and thus we must consider what level of capability would be required to perform each attack vector in order to properly model risk and prioritize mitigations. Since many of the attacks are already performed in the real-world by real actors, the ability required to perform the attack could also be a factor in the tradeoff.

\textbf{Capability Classification} We classify the capabilities of the attacker for each attack vector across three dimensions: privilege level (user, admin/root, physical access, or supply chain/manufacturer), technical sophistication (low -- tools for the attack are easily available with no additional knowledge needed; medium -- attack requires some knowledge or custom tooling; or high -- requires knowledge comparable to a nation-state), and resource requirement (software, hardware like a single board computer or USB adapter, or expensive hardware like an oscilloscope, RF lab, or microscope). This is summarized in Table~\ref{tab:capability-analysis}.

\section{Conclusion}
\label{sec:conclusion}

To understand the true nature of the assumed security properties of FIDO2, we have dissected the FIDO2/WebAuthn trade model~\cite{dbase:fido2-formal-analysis,localattacks2023} and identified 8 attack vectors that can compromise those properties - malicious web extensions, platform handler malware, passive wiretapping, virtual devices, direct CTAP malware, USB/HID hardware implants, malicious USB hubs/docks/extenders, and NFC relay attacks - which reveal that the security properties of FIDO2 depend heavily on assumptions that may not always be valid. Specifically, we show that: Metadata leaks from side channels can not only be used to steal credentials, but can also be used to extract AAGUIDs and fingerprint users based on their authentication activities~\cite{mojoauth2026aaguid,yubico_attestation}. The FIDO2 trade model assumes a trusted surrounding environment, but ultimately malware can subvert this assumption and gain full control of the authenticator's functionality. Phishing resistance is not enough to protect users if the attacker is able to intercept the data before and after the crypto phase~\cite{squarex2025passkeyspwned}.

We further show how these attack vectors can be chained together to enable even stronger attacks, stressing the importance of designing trade solutions that defend against multiple layers of vulnerability at once.

In conclusion, this paper demonstrates that there are means to subvert or circumvent many of the assumed security properties of FIDO2. Although the FIDO2 standard is still much better than passwords, it is critical to know exactly what it protects against in order to effectively defend against such types of attacks. Going forward, research should be focused on discovering methods for mitigating these vulnerabilities in order to make sure that FIDO2 actually delivers on the advantages its advocates anticipate it will deliver.


\section*{Acknowledgments}
This research was conducted as part of the Research and Development (R\&D) efforts at DigitalFortress Private Limited \& Indominus Labs Private Limited. We acknowledge the company's support in facilitating this study. This work is protected and is not intended for reuse in any commercial capacity.

\bibliographystyle{IEEEtran}
\bibliography{reference.bib}

@inproceedings{dbase:fido2-usability,
  author    = {Ghorbani Lyastani, Sanam and Schilling, Michael and Neumayr, Michaela and Backes, Michael and Bugiel, Sven},
  title     = {Is {FIDO2} the Kingslayer of User Authentication? {A} Comparative Usability Study of {FIDO2} Passwordless Authentication},
  booktitle = {2020 IEEE Symposium on Security and Privacy (S\&P)},
  pages     = {268--285},
  year      = {2020},
  month     = {5},
  publisher = {IEEE}
}

@inproceedings{dbase:fido2-formal-analysis,
  author    = {Barbosa, Manuel and Boldyreva, Alexandra and Chen, Shan and Warinschi, Bogdan},
  title     = {Provable Security Analysis of {FIDO2}},
  booktitle = {Advances in Cryptology -- CRYPTO 2021},
  pages     = {125--156},
  year      = {2021},
  publisher = {Springer International Publishing}
}

@article{dbase:fido2-enterprise-challenges,
  author  = {Kepkowski, Michal and Machulak, Maciej and Wood, Ian and Kaafar, Dali},
  title   = {Challenges with Passwordless {FIDO2} in an Enterprise Setting: {A} Usability Study},
  journal = {arXiv preprint arXiv:2308.08096},
  year    = {2023}
}

@misc{bindel2022ctap21,
  author       = {Bindel, Nina and Cremers, Cas and Zhao, Mang},
  title        = {{FIDO2}, {CTAP} 2.1, and {WebAuthn} 2: Provable Security and Post-Quantum Instantiation},
  howpublished = {Cryptology ePrint Archive, Paper 2022/1029},
  year         = {2022},
  url          = {https://eprint.iacr.org/2022/1029}
}

@misc{ravilla2024study,
  author       = {Ravilla, Harshavardhan and others},
  title        = {Study and Analysis of {FIDO2} Passwordless Web Authentication},
  howpublished = {ResearchGate preprint},
  year         = {2024},
  month        = {8},
  url          = {https://www.researchgate.net/publication/383944347_Study_and_Analysis_of_FIDO2_Passwordless_Web_Authentication}
}

@misc{localattacks2023,
  author       = {Bakalis, Odysseas and Menges, Florian and others},
  title        = {A Security and Usability Analysis of Local Attacks Against {FIDO2}},
  howpublished = {arXiv preprint arXiv:2308.02973},
  year         = {2023},
  url          = {https://arxiv.org/pdf/2308.02973}
}

@techreport{ninjalab2024eucleak,
  author      = {Roche, Thomas},
  title       = {{EUCLEAK} Side-Channel Attack on the {YubiKey} 5 Series (Revealing and Breaking Infineon {ECDSA} Implementation on the Way)},
  institution = {NinjaLab},
  year        = {2024},
  month       = {9},
  url         = {https://ninjalab.io/wp-content/uploads/2024/09/20240903_eucleak.pdf}
}

@inproceedings{roche2024eucleakieee,
  author    = {Roche, Thomas},
  title     = {{EUCLEAK} Side-Channel Attack on the {YubiKey} 5 Series},
  booktitle = {IEEE Conference Publication},
  year      = {2025},
  publisher = {IEEE},
  url       = {https://ieeexplore.ieee.org/document/11023335/}
}

@misc{w3cwebauthn2021,
  author       = {{W3C Web Authentication Working Group}},
  title        = {Web Authentication: An {API} for Accessing Public Key Credentials Level 2},
  howpublished = {W3C Recommendation},
  year         = {2021},
  url          = {https://www.w3.org/TR/webauthn-2/}
}

@misc{fidoctap2,
  author       = {{FIDO Alliance}},
  title        = {Client to Authenticator Protocol ({CTAP}), {CTAP2} Specification},
  howpublished = {FIDO Alliance Proposed Standard},
  year         = {2019},
  url          = {https://fidoalliance.org/specs/fido-v2.0-ps-20190130/fido-client-to-authenticator-protocol-v2.0-ps-20190130.html}
}

@misc{fidomds2021,
  author       = {{FIDO Alliance}},
  title        = {{FIDO} Metadata Statement, v3.0},
  howpublished = {FIDO Alliance Proposed Standard},
  year         = {2021},
  month        = {5},
  url          = {http://fidoalliance.org/specs/mds/fido-metadata-statement-v3.0-ps-20210518.html}
}

@misc{squarex2025passkeyspwned,
  author       = {{SquareX Labs}},
  title        = {Passkeys Pwned: Turning {WebAuthn} Against Itself},
  howpublished = {SquareX Labs Blog},
  year         = {2025},
  month        = {8},
  url          = {https://labs.sqrx.com/passkeys-pwned-turning-webauth-against-itself-0dbddb7ade1a}
}

@misc{hackernews2025bypasssynced,
  author       = {{The Hacker News}},
  title        = {How Attackers Bypass Synced Passkeys},
  howpublished = {The Hacker News},
  year         = {2025},
  month        = {10},
  url          = {https://thehackernews.com/2025/10/how-attackers-bypass-synced-passkeys.html}
}

@misc{francis2011relay,
  author       = {Francis, Lishoy and Hancke, Gerhard and Mayes, Keith and Markantonakis, Konstantinos},
  title        = {Practical Relay Attack on Contactless Transactions by Using {NFC} Mobile Phones},
  howpublished = {Cryptology ePrint Archive, Paper 2011/618},
  year         = {2011},
  url          = {https://eprint.iacr.org/2011/618.pdf}
}

@misc{supercardx2025,
  author       = {{The Hacker News}},
  title        = {{SuperCard X} {Android} Malware Enables Contactless {ATM} and {PoS} Fraud via {NFC} Relay Attacks},
  howpublished = {The Hacker News},
  year         = {2025},
  month        = {4},
  url          = {https://thehackernews.com/2025/04/supercard-x-android-malware-enables.html}
}

@misc{hak5omg,
  author       = {{Hak5}},
  title        = {{O.MG} Cable Product Documentation},
  howpublished = {Hak5 Shop},
  url          = {https://shop.hak5.org/products/omg-cable}
}

@misc{vice2021omg,
  author       = {{Vice / Motherboard}},
  title        = {This Seemingly Normal Lightning Cable Will Leak Everything You Type},
  howpublished = {Vice},
  year         = {2021},
  url          = {https://www.vice.com/en/article/omg-cables-keylogger-usbc-lightning/}
}

@misc{darkreading2023implants,
  author       = {{Dark Reading}},
  title        = {From `{O.MG}' to {NSA}, What Hardware Implants Mean for Security},
  howpublished = {Dark Reading},
  year         = {2023},
  month        = {12},
  url          = {https://www.darkreading.com/threat-intelligence/from-o-mg-to-nsa-what-hardware-implants-mean-for-security}
}

@misc{mgmspecops2023,
  author       = {{Specops Software}},
  title        = {{MGM} Resorts Service Desk Hack},
  howpublished = {Specops Software Blog},
  year         = {2023},
  url          = {https://specopssoft.com/blog/mgm-resorts-service-desk-hack/}
}

@misc{yubico_attestation,
  author       = {{Yubico Developer Program}},
  title        = {{WebAuthn} Attestation and Authenticator Metadata},
  howpublished = {Yubico Developer Documentation},
  url          = {https://developers.yubico.com/Developer_Program/WebAuthn_Starter_Kit/Attestation.html}
}

@misc{mojoauth2026aaguid,
  author       = {{MojoAuth Blog}},
  title        = {Passkey {AAGUID}s: Which Password Manager Is Your User Actually Using?},
  howpublished = {MojoAuth Blog},
  year         = {2026},
  url          = {https://mojoauth.com/blog/passkey-aaguid-password-manager-mapping}
}

@misc{w3c_webextensions_issue361,
  author       = {{W3C WebExtensions Community Group}},
  title        = {Issue \#361: Interaction between Web Extensions and the Credential Management / {WebAuthn} {APIs}},
  howpublished = {GitHub Issue},
  url          = {https://github.com/w3c/webextensions/issues/361}
}

@misc{vsmartcard_morgner,
  author       = {Morgner, Frank},
  title        = {Virtual Smart Card / Remote Reader Project (vsmartcard)},
  howpublished = {Project Website},
  url          = {https://frankmorgner.github.io/vsmartcard/}
}

@misc{fido_security_bulletin,
  author       = {Mitra, Aditya},
  title        = {{FIDO} Alliance Security Bulletin 1721 (Draft), {NFC} Relay Attack Disclosure},
  year         = {2023},
  url          = {https://adityamitra5102.github.io/certs/FIDO-SecurityBulletin1721_DRAFT.pdf}
}

@misc{fido2vectors2026,
  author       = {{arXiv preprint}},
  title        = {An Analysis of Attack Vectors Against {FIDO2} Authentication},
  howpublished = {arXiv preprint arXiv:2604.20826},
  year         = {2026},
  url          = {https://arxiv.org/html/2604.20826v1}
}

@misc{cve202445678eucleak,
  author       = {{MITRE}},
  title        = {{CVE-2024-45678}: {Y}ubico {Y}ubiKey 5 Series and {Y}ubiHSM 2 {ECDSA} Secret-Key Extraction ({EUCLEAK})},
  howpublished = {National Vulnerability Database},
  year         = {2024},
  url          = {https://nvd.nist.gov/vuln/detail/CVE-2024-45678}
}

@misc{fido_security_ref,
  author       = {{FIDO Alliance}},
  title        = {{FIDO} Security Reference},
  howpublished = {FIDO Alliance Review Draft},
  year         = {2021},
  month        = may,
  url          = {https://fidoalliance.org/specs/common-specs/fido-security-ref-v2.1-rd-20210525.html},
  note         = {Accessed: 2026-08-09}
}

@misc{mollema2025actortokens,
  author       = {Mollema, Dirk-jan},
  title        = {One Token to Rule Them All: Obtaining Global Admin in Every {Entra ID} Tenant via Actor Tokens},
  howpublished = {dirkjanm.io Blog; presented at Black Hat USA and {DEF CON} 33},
  year         = {2025},
  month        = sep,
  url          = {https://dirkjanm.io/obtaining-global-admin-in-every-entra-id-tenant-with-actor-tokens/},
  note         = {CVE-2025-55241}
}

@misc{thehackernews2026winhello,
  author       = {Khandelwal, Swati},
  title        = {Malware Can Abuse {Windows Hello} for Business Keys for Persistent {Entra ID} Access},
  howpublished = {The Hacker News},
  year         = {2026},
  month        = aug,
  url          = {https://thehackernews.com/2026/08/malware-can-abuse-windows-hello-for.html},
  note         = {Reporting on research by Dirk-jan Mollema}
}

\end{document}